\documentclass[final,3p,times]{elsarticle}
\usepackage{amsfonts}
\usepackage{subfigure}
\usepackage{amssymb}
\usepackage{amsmath}
\usepackage{graphics}
\usepackage{epsfig}
\usepackage{epstopdf}
\usepackage{amsthm}
\usepackage{textcomp}
\usepackage{color}
\usepackage{float}
\usepackage{multirow}

\usepackage{algorithmic}
\usepackage{algorithm}

\journal{Applied Mathematical Modelling}

\begin{document}

\begin{frontmatter}

\title{On the effectiveness of the truth-spreading/rumor-blocking strategy \\ for restraining rumors}

\cortext[cor1]{Corresponding author}

\author[rvt,rvt2]{Lu-Xing Yang}
\ead{ylx910920@gmail.com}

\author[rvt3]{Tianrui Zhang}
\ead{363726657@qq.com}

\author[rvt3]{Xiaofan Yang\corref{cor1}}
\ead{xfyang1964@gmail.com}

\author[rvt3]{Yingbo Wu}
\ead{wyb@cqu.edu.cn}

\author[rvt4]{Yuan Yan Tang}
\ead{yytang@umac.mo}

\address[rvt]{School of Mathematics and Statistics, Chongqing University, Chongqing, 400044, China}

\address[rvt2]{Faculty of Electrical Engineering, Mathematics and Computer Science, Delft University of Technology,  Delft, GA 2600, The Netherlands}

\address[rvt3]{School of Software Engineering, Chongqing University, Chongqing, 400044, China}

\address[rvt4]{Department of Computer and Information Science, The University of Macau, Macau}

\begin{abstract}
Spreading truths and blocking rumors are two typical strategies for inhibiting rumors. In practice, a tradeoff between the two strategies, which is known as the TSRB strategy, may achieve a better cost-effectiveness. This paper is devoted to assessing the effectiveness of the TSRB strategy. For that purpose, an individual-level spreading model (the generic URQT model) capturing the interaction between a rumor and the truth is established. Under the model, a set of criteria for the dying out of a rumor is presented. These criteria capture the combined influence of the basic parameters and the network structures on the effectiveness of the TSRB strategy. Experimental results show that, when the rumor dies out, the dynamics of a simplified URQT model (the linear URQT model) fits well with the actual rumor-truth interacting process. Therefore, the generic URQT model and sometimes the linear URQT model provide a proper basis for assessing the effectiveness of the TSRB strategy.
\end{abstract}


\end{frontmatter}



\section{Introduction}

Rumors are an important form of interpersonal communications, and rumor spreading has a significant impact on human affairs. The emerging online social networks (OSNs) offer a shortcut for the fast spread of rumors, immensely expanding the influence of rumors \cite{Viswanath2009, Kwak2010, Doerr2012}. Unfortunately, most rumors could induce social panic or/and economic loss \cite{Thomas2007}. For instance, the rumor of "explosions at White House had injured Obama" dispersed by Syrian hackers through the invaded twitter account of Associated Press lead to 10 billion USD losses in only a few minutes \cite{Peter2013}. Consequently, the issue of containing the diffusion of rumors through OSNs has become a hotspot of research in the field of cybersecurity \cite{Budak2011}.

The mission of rumor spreading dynamics is to model and study the spreading process of rumors, so as to understand the influence of different factors on rumor spreading and thereby to develop cost-effective rumor-containing strategies. Since the seminal work by Daley and Kendall \cite{Daley1964}, a batch of rumor spreading models based on homogeneous networks have been proposed, aiming at understanding the influence of the basic parameters on the prevalence of rumors \cite{ZhaoLJ2011a, ZhaoLJ2012a, HuoLA2012a, Afassinou2014, ZhaoLJ2015, HuoLA2015, HuoLA2016}. Emperical studies starting at the end of last century fully show that OSNs are heterogeneous rather than homogeneous \cite{Albert2002, Ebel2002}. From then on, a large body of rumor spreading models based on complex networks have been suggested, with emphasis on the combined influence of the basic parameters and the network structures on rumor spreading \cite{Nekovee2007, ZhouJ2007, Roshani2012, ZhaoLJ2013a, WangYQ2013, Naimi2013, ZanYL2014, HeZB2015, HeZB2017, LiuQM2017, QiuXY2016, Oliveros2017}. As these models are established by use of the crude mean-field theory, their dynamics may severely deviate from the actual rumor spreading process, and the actual influence of the network structures on rumor spreading cannot be made clear by studying these models.

The continuous-time individual-level modeling technique developed by Van Mieghem et al. \cite{Mieghem2009} has been recognized as a powerful tool to understand epidemic processes, because the model established using the technique accurately captures the average dynamics of a class of epidemics and is successfully applied to probing the impact of the network structure on the epidemic spreading \cite{Mieghem2009, Mieghem2011}. Exploiting the modeling technique, a group of individual-level epidemic models have been reported consecutively \cite{Sahneh2012, Sahneh2013, YangLX2015, XuSH2015, ZhengR2015, YangLX2017a, YangLX2017b}. In our opinion, the individual-level modeling technique is especially suited to the study of rumor spreading. To our knowledge, to date the literature in this aspect is still rare.

Spreading truths and blocking rumors are two typical strategies for inhibiting rumors in OSNs \cite{WenS2014, WenS2015}. However, everything has pros and cons. On one hand, a rumor can be clarified by circulating the truth in OSNs, at the cost of acquiring convincing evidence supporting the truth. On the other hand, the negative influence of a rumor can be diminished by quarantining a group of rumor spreaders, at the risk of violating human rights. In practice, a tradeoff between the two rumor-containing strategies, which is known as the truth-spreading/rumor-blocking (TSRB) strategy, may achieve a better cost-effectiveness.

This paper aims to assess the effectiveness of the TSRB strategy. For that purpose, a continuous-time individual-level rumor-truth interacting model (the generic URQT model) is derived. Then a set of criteria for the dying out of a rumor is presented. These criteria captures the combined influence of the basic parameters and the network structures on the effectiveness of the truth-spreading strategy. Simulation experiments show that, when the rumor dies out, the dynamics of a simplified URQT model (the linear URQT model) fits well with the actual rumor-truth interacting process. Therefore, the generic URQT model and sometimes the linear URQT model provide a proper basis for assessing the effectiveness of the TSRB strategy.

The subsequent materials are organized in this fashion. Section 2 derives the original URQT model, the exact URQT model, the linear URQT model, and the generic URQT model, sequentially. Section 3 studies the dynamics of the generic URQT model. Simulation experiments examining the effectiveness of the linear URQT model are conducted in Section 4. Finally, Section 5 summarizes this work.

\newtheorem{rk}{Remark}
\newproof{pf}{Proof}
\newtheorem{thm}{Theorem}
\newtheorem{lm}{Lemma}
\newtheorem{exm}{Example}
\newtheorem{cor}{Corollary}
\newtheorem{de}{Definition}
\newtheorem{cl}{Claim}
\newtheorem{pro}{Proposition}
\newtheorem{con}{Conjecture}

\newtheorem{app}{Appendix}

\newproof{pfle2}{Proof of Lemma 2}
\newproof{pfth1}{Proof of Theorem 1}
\newproof{pfth4}{Proof of Theorem 4}
\newproof{pfcl1}{Proof of Claim 1}

\section{The formation of the generic URQT model}

This section is devoted to establishing a continuous-time individual-level dynamic model capturing the interaction between a rumor and the truth.

\subsection{Notions, notations and basic hypotheses}

Consider an online social network (OSN) consisting of $N$ persons labelled $1, 2, \cdots, N$. Let $V = \{1, 2, \cdots, N\}$. Suppose a rumor spreads in the OSN through a rumor-spreading network $G_R = (V, E_R)$, where $(i,j) \in E_R$ if and only if person $j$ can tell the rumor to person $i$. Suppose the truth against the rumor spreads in the OSN through a truth-spreading network $G_T = (V, E_T)$, where $(i,j) \in E_T$ if and only if person $j$ can tell the truth to person $i$. Hereafter, the two networks are always assumed to be strongly connected.

At any time after the appearance of the rumor and the truth, every person in the OSN is assumed to be in one of four possible states: \emph{uncertain}, \emph{rumor-spreading}, \emph{quarantined}, and \emph{truth-believing}. An uncertain person believes neither the rumor nor the truth. A rumor-spreader believes the rumor and is allowed to spread the rumor. A quarantined person believes the rumor but is not allowed to spread the rumor. A truth-believer believes the truth (and is allowed to spread the truth). Depending on the personal judgment or/and knowledge, a person may choose to be uncertain or to believe the rumor or to believe the truth. A rumor-believer is likely to be quarantined, and the likelihood is proportional to his influence in the OSN. Let $X_i(t)$ = 0, 1, 2, and 3 denote that at time $t$ person $i$ is uncertain, rumor-spreading, quarantined, and truth-believing, respectively. The state of the OSN at time $t$ is represented by the vector
\[
\mathbf{X}(t) = (X_1(t), X_2(t), \cdots, X_N(t))^T.
\]

Next, let us introduce a basic set of hypotheses as follows.

\begin{enumerate}
	
	\item[(H$_1$)] Due to the influence of a rumor-spreader $j$, at any time an uncertain person $i$ becomes rumor-spreading at rate $\beta_{ij}^U \geq 0$. Here, $\beta_{ij}^U > 0$ if and only if $(i,j) \in E_R$. Let $\mathbf{B}_U = \left[\beta_{ij}^U\right]_{N \times N}$. This hypothesis captures the influence of rumor-spreaders on uncertain persons.
	
	\item[(H$_2$)] Due to the influence of a rumor-spreader $j$, at any time a truth-believer $i$ becomes rumor-spreading at rate $\beta_{ij}^T$. Here, $\beta_{ij}^T > 0$ if and only if $(i,j) \in E_R$. Let $\mathbf{B}_T= \left[\beta_{ij}^T\right]_{N \times N}$. This hypothesis captures the influence of rumor-spreaders on truth-believers.
	
	\item[(H$_3$)] Due to the influence of a truth-believer $j$, at any time an uncertain person $i$ becomes truth-believing at rate $\gamma_{ij}^U$. Here, $\gamma_{ij}^U > 0$ if and only if $(i,j) \in E_T$. This hypothesis captures the influence of truth-believers on uncertain persons.
	
	\item[(H$_4$)] Due to the influence of a truth-believer $j$, at any time a rumor-spreader $i$ becomes truth-believing at rate $\gamma_{ij}^R$. Here, $\gamma_{ij}^R > 0$ if and only if $(i,j) \in E_T$. This hypothesis captures the influence of truth-believers on rumor-spreaders.
	
	\item[(H$_5$)] At any time a rumor-spreader $i$ becomes quarantined at rate $\theta_i > 0$. Let $\mathbf{D}_{\theta} = diag\left(\theta_i\right)$.
	
	\item[(H$_6$)] Due to the education, at any time a quarantined person $i$ becomes truth-believing at rate $\delta_i > 0$.  Let $\mathbf{D}_{\delta} = diag\left(\delta_i\right)$.
	
\end{enumerate}

All the forthcoming models are assumed to comply with these basic hypotheses.

\subsection{The original URQT model}

For fundamental knowledge on continuous-time Markov chain, see Ref. \cite{Stewart2009}.

Another way of representing the OSN state at time $t$ is by the decimal number
\[
i(t)=\sum_{k=1}^{N}X_k(t)4^{k-1}.
\]
In this context, there are totally $4^N$ possible OSN states: $0, 1, \cdots, 4^{N} - 1$. According to the basic hypotheses, the infinitesimal generator $\mathbf{Q}=\left[q_{ij}\right]_{4^N \times 4^N}$ for the rumor-truth interacting process is given as
	\begin{equation}
	q_{ij}=\left\{
	\begin{aligned}
	&\theta_m   & \text{if} &\quad i=j-4^{m-1}; m=1,2,\cdots,N, x_{m}=1 \\
	&\delta_m  & \text{if} &\quad i=j-4^{m-1}; m=1,2,\cdots,N, x_{m}=2 \\
	&\sum_{k=1}^{N}\beta_{mk}^U1_{\{x_{k}=1\}}   & \text{if} &\quad i=j-4^{m-1}; m=1,2,\cdots,N, x_{m}=0  \\
	&\sum_{k=1}^{N}\beta_{mk}^T1_{\{x_{k}=1\}}   & \text{if} &\quad i=j+2\cdot4^{m-1}; m=1,2,\cdots,N, x_{m}=3  \\
	&\sum_{k=1}^{N}\gamma_{mk}^U1_{\{x_{k}=3\}}  & \text{if}&
	\quad i=j-3\cdot4^{m-1}; m=1,2,\cdots,N, x_{m}=0 \\
	&\sum_{k=1}^{N}\gamma_{mk}^R1_{\{x_{k}=3\}}  & \text{if}&
	\quad i=j-2\cdot4^{m-1}; m=1,2,\cdots,N, x_{m}=1 \\
	&-\sum_{k=0,k\neq i}^{N-1}q_{ik}  & \text{if} & \quad i=j\\
	& 0 \quad  & \quad  &\text{otherwise}.
	\end{aligned}
	\right.
	\end{equation}
where $i=\sum_{k=1}^{N}x_{k}4^{k-1}$, $1_A$ stands for the indicator function of set $A$. The continuous-time Markov chain with the infinitesimal generator $\mathbf{Q}$, which accommodates an infinite number of sample paths, accurately captures the dynamics of the interaction between the rumor and the truth. We refer to the Markov chain model as the \emph{original uncertain-rumor-quarantine-truth (URQT) model}. See Fig. 1 for the state transition rates of person $i$ under the model.

\begin{figure}[!t]
	\centering
	\includegraphics[width=2.5in]{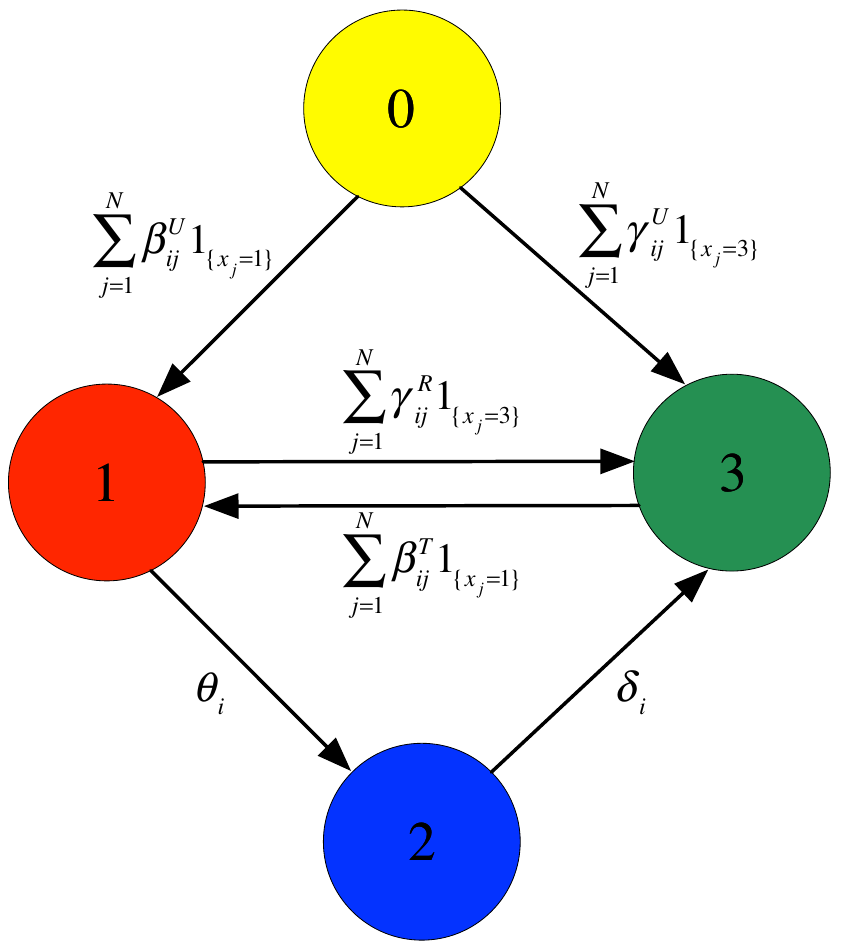}
	\caption{The state transition rates of person $i$ under the original URQT model.}
\end{figure}

\subsection{The exact URQT model}

Let $s_i(t)$ denote the probability that at time $t$ the OSN state is $i = \sum_{k=1}^{N}x_{k}4^{k-1}$. That is,
\[
s_{i}(t)=\Pr\left\{X_{1}(t)=x_{1},\cdots,X_{N}(t)=x_N\right\}.
\]
Then, the vector-valued function $\mathbf{s}(t)=[s_{0}(t),\cdots,s_{4^{N}-1}(t)]^T$ obeys the linear differential system
\begin{equation}
\frac{d\mathbf{s}^{T}(t)}{dt}=\mathbf{s}^{T}(t)\mathbf{Q}.
\end{equation}
\noindent This model accurately captures the average dynamics of the rumor-truth interaction. Therefore, we refer to the model as the \emph{exact URQT model}. Unfortunately, the state transition rates of a person under the model cannot be graphically illustrated. Although the exact URQT model is a linear differential system, its dimensionality grows exponentially with the increasing size of the OSN, leading to mathematical intractability.

To give another form of the exact URQT model, let us introduce the following notations.
\[U_i(t) = \Pr\{X_i(t) = 0\}, \quad R_i(t) = \Pr\{X_i(t) = 1\}, \quad Q_i(t) = \Pr\{X_i(t) = 2\}, \quad T_i(t) = \Pr\{X_i(t) = 3\}.
\]
As $U_i(t) + R_i(t) + Q_i(t) + T_i(t) = 1$, the vector $(U_i(t), R_i(t), T_i(t))$ captures the average state of person $i$ at time $t$. The following lemma gives an equivalent form of the exact URQT model.

\begin{lm}
	The exact URQT model is equivalent to
		\begin{equation}
		\left\{
		\begin{aligned}
		\frac{dU_i(t)}{dt} &= -\sum_{j = 1}^N \beta_{ij}^U \Pr\{X_i(t) = 0, X_j(t) = 1\}-\sum_{j = 1}^N \gamma_{ij}^U \Pr\{X_i(t) = 0, X_j(t) = 3\}, \\
		\frac{dR_i(t)}{dt} &= \sum_{j = 1}^N \beta_{ij}^U \Pr\{X_i(t) = 0, X_j(t) = 1\}+\sum_{j = 1}^N \beta_{ij}^T \Pr\{X_i(t) = 3, X_j(t) = 1\} \\
		&\quad-\sum_{j = 1}^N \gamma_{ij}^R \Pr\{X_i(t) = 1, X_j(t) = 3\} - \theta_i R_i(t), \\
		\frac{dT_i(t)}{dt} &= \sum_{j = 1}^N \gamma_{ij}^U \Pr\{X_i(t) = 0, X_j(t) = 3\}+\sum_{j = 1}^N \gamma_{ij}^R \Pr\{X_i(t) = 1, X_j(t) = 3\} \\
		&\quad-\sum_{j = 1}^N \beta_{ij}^T \Pr\{X_i(t) = 3, X_j(t) = 1\}+ \delta_i(1-U_i(t)-R_i(t)-T_i(t)), \\
		& \quad \quad t \geq 0, i = 1, 2, \cdots, N.
		\end{aligned}
		\right.
		\end{equation}
\end{lm}

The proof of this lemma is left to Appendix A. The equivalent model is not closed and hence cannot be studied directly.

\subsection{The linear URQT model}

In order to make the exact URQT model easily treatable, let us make an added
set of hypotheses as follows.

\begin{enumerate}
	\item[(H$_7$)] $\Pr\{X_i(t) = 0, X_j(t) = 1\} = U_i(t)R_j(t), \quad 1 \leq i, j \leq N, i \neq j.$
	\item[(H$_8$)] $\Pr\{X_i(t) = 0, X_j(t) = 3\} = U_i(t)T_j(t),  \quad 1 \leq i, j \leq N, i \neq j.$
	\item[(H$_9$)] $\Pr\{X_i(t) = 1, X_j(t) = 3\} = R_i(t) T_j(t), \quad 1 \leq i, j \leq N, i \neq j$.
	\item[(H$_{10}$)]  $\Pr\{X_i(t) = 3, X_j(t) = 1\} = T_i(t) R_j(t), \quad 1 \leq i, j \leq N, i \neq j $.
\end{enumerate}

\noindent Based on the equivalent model (3) and these independence hypotheses, we derive the following model.
	\begin{equation}
	\left\{
	\begin{aligned}
	\frac{dU_i(t)}{dt} &= -U_i(t)\sum_{j = 1}^N \beta_{ij}^U R_j(t)-U_i(t)\sum_{j = 1}^N \gamma_{ij}^U T_j(t)\\
	\frac{dR_i(t)}{dt} &= U_i(t)\sum_{j = 1}^N \beta_{ij}^U R_j(t)+T_i(t)\sum_{j = 1}^N \beta_{ij}^T R_j(t)-R_i(t)\sum_{j = 1}^N\gamma_{ij}^R T_j(t)- \theta_i R_i(t)\\
	\frac{dT_i(t)}{dt} &= U_i(t)\sum_{j = 1}^N \gamma_{ij}^U T_j(t)+R_i(t)\sum_{j = 1}^N \gamma_{ij}^R T_j(t)-T_i(t)\sum_{j = 1}^N \beta_{ij}^T R_j(t)+\delta_i(1-U_i(t)-R_i(t)-T_i(t)), \\
	& \quad \quad t \geq 0, i = 1, 2, \cdots, N.
	\end{aligned}
	\right.
	\end{equation}

\noindent We refer to the model as the \emph{linear URQT model}, because the spreading rates are each linear in its arguments. See Fig. 2 for the state transition rates of person $i$ under the linear URQT model.

\begin{figure}[!t]
	\centering
	\includegraphics[width=2.5in]{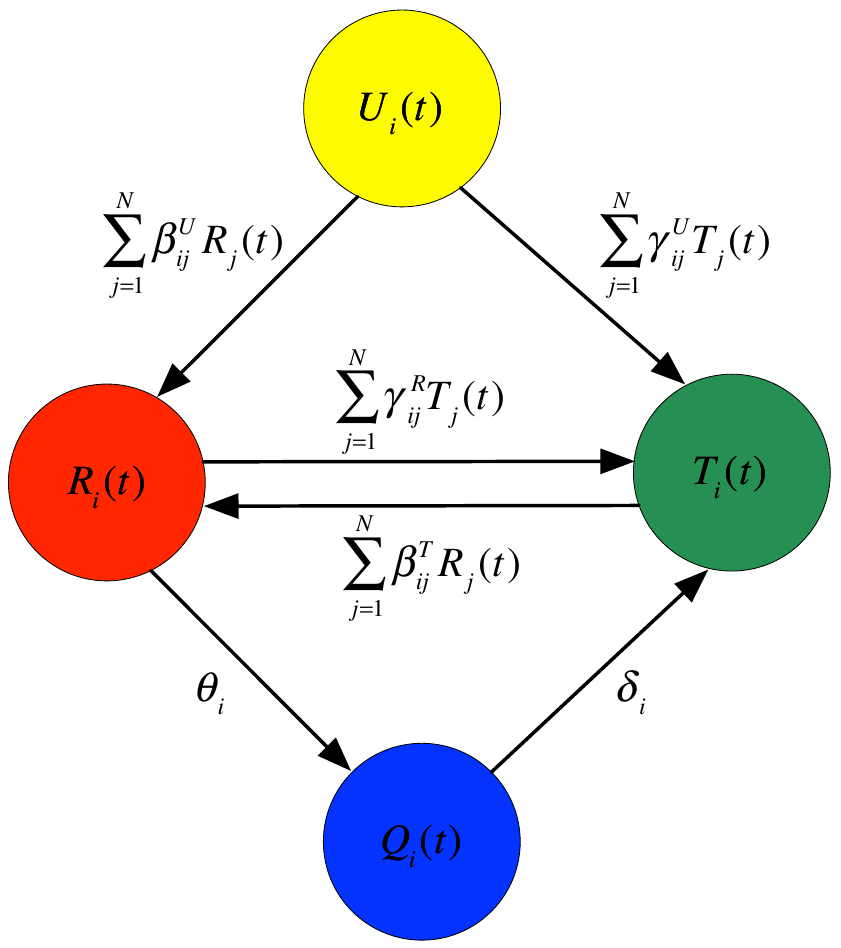}
	\caption{The state transition rates of person $i$ under the linear URQT model.}
\end{figure}

The linear URQT model is a 3$N$-dimensional dynamical system and hence is mathematically tractable. However, due to the difference between the linear spreading rates and the actual spreading rates, the dynamics of the model may deviate from the actual average dynamics of the rumor-truth interaction.

\subsection{The generic URQT model}

In practical applications, the independence assumptions given in the previous subsection may not hold. If this is the case, the linear URQT model would fail to accurately capture the average dynamics of the rumor-truth interaction. For the purpose of approximating the exact URQT model more accurately, let us introduce a more general rumor-truth interacting model as follows.
	\begin{equation}
	\left\{
	\begin{aligned}
	\frac{dU_i(t)}{dt} &= -U_i(t)f_i^U(R_1(t),\cdots,R_N(t))-U_i(t)g_i^U(T_1(t),\cdots,T_N(t))\\
	\frac{dR_i(t)}{dt}&=U_i(t)f_i^U\left(R_1(t), \cdots, R_N(t)\right)+T_i(t)f_i^T\left(R_1(t), \cdots, R_N(t)\right)-R_i(t)g_i^R(T_1(t),\cdots,T_N(t))-\theta_iR_i(t),\\
	\frac{dT_i(t)}{dt}&=U_i(t)g_i^U\left(T_1(t), \cdots, T_N(t)\right)+R_i(t)g_i^R\left(T_1(t), \cdots, T_N(t)\right) -T_i(t)f_i^T(R_1(t),\cdots,R_N(t))\\
	&\quad + \delta_i(1-U_i(t)-R_i(t)-T_i(t)), \\
	&\quad \quad t \geq 0, i=1,2,\cdots, N.
	\end{aligned}
	\right.
	\end{equation}
Here, the spreading rates, $f_i^U$, $f_i^T$, $g_i^U$ and $g_i^R$, satisfy the following generic conditions.

\begin{enumerate}
	
	\item[(C$_1$)] (Proximity) An uncertain person or a truth-believer can and can only be influenced by rumor-spreaders who can tell the rumor to the person. That is, $f_i^U$ and $f_i^T$ are dependent upon $R_j(t)$ if and only if $(i, j) \in E_R$. Likewise, An uncertain person or a rumor-spreader can and can only be influenced by truth-believers who can tell the truth to the person. That is, $g_i^U$ and $g_i^R$ is dependent upon $T_j(t)$ if and only if $(i,j) \in E_T$.
	
	\item[(C$_2$)] (Nullity) The rumor cannot spread in the OSN unless there is a rumor-spreader in the OSN. That is, $f_i^U(0, \cdots, 0)=f_i^T(0, \cdots, 0)=0$. Likewise, the truth cannot spread in the OSN unless there is a truth-believer in the OSN. That is, $g_i^U(0, \cdots, 0)=g_i^R(0, \cdots, 0)=0$.
	
	\item[(C$_3$)] (Smoothness) The spreading rates are twice continuously differentible.
	
	\item[(C$_4$)] (Monotonicity) The spreading rates are each strictly increasing in its arguments.
	
	\item[(C$_5$)] (Concavity) The spreading rates are each concave in its arguments.
	
\end{enumerate}

We refer to the model as the \emph{generic URQT model}. See Fig. 3 for the state transition rates of person $i$ under the model.

\begin{figure}[!t]
	\centering
	\includegraphics[width=2.5in]{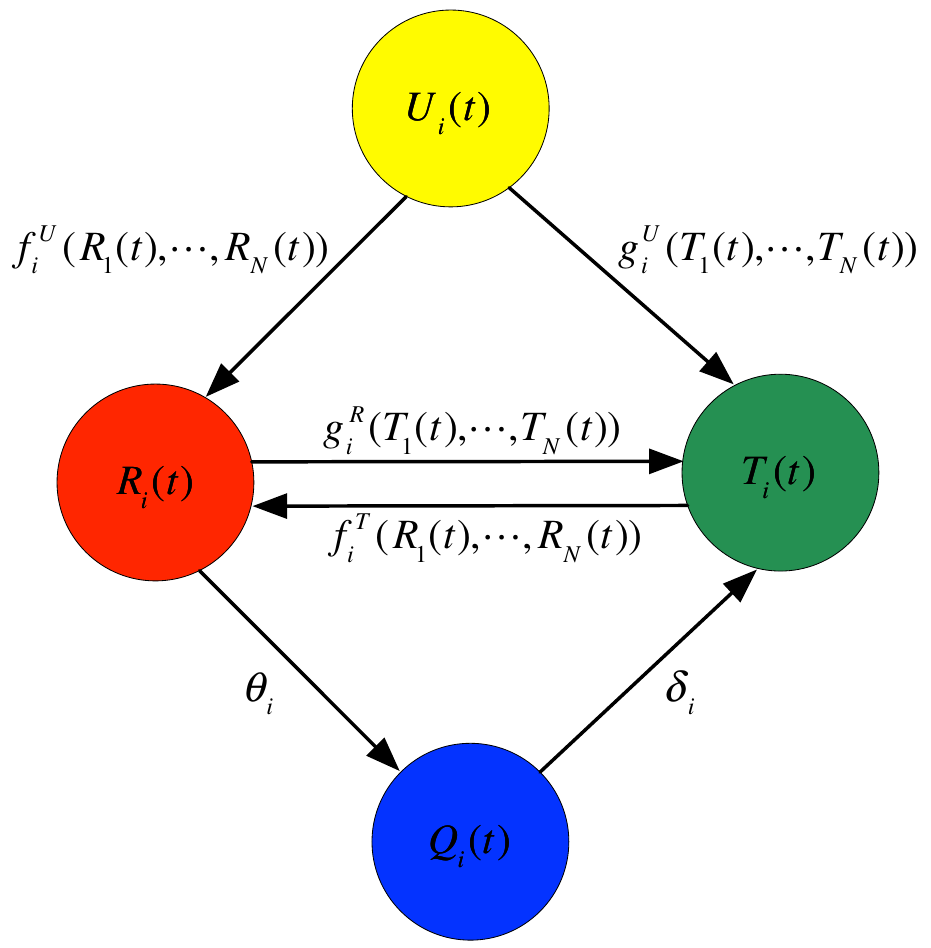}
	\caption{The state transition rates of person $i$ under the generic URQT model.}
\end{figure}

The generic URQT model subsumes the linear URQT model as well as many URQT models with specific nonlinear spreading rates. By properly choosing the generic spreading rates, the generic URQT model can approximate the exact URQT model better than the linear URQT model.

\subsection{The generic SURQT model}

In the case that all persons in the OSN are not uncertain, we have $\lim_{t\rightarrow}U_i(t)=0$ for $1\leq i\leq N$. Consider the following limit system of model (5).
	\begin{equation}
	\left\{
	\begin{aligned}
	\frac{dR_i(t)}{dt}&=T_i(t)f_i^T\left(R_1(t), \cdots, R_N(t)\right)-R_i(t)g_i^R(T_1(t),\cdots,T_N(t)) -\theta_iR_i(t),\\
	\frac{dT_i(t)}{dt}&=R_i(t)g_i^R\left(T_1(t), \cdots, T_N(t)\right)-T_i(t)f_i^T(R_1(t),\cdots,R_N(t)) +\delta_i(1-R_i(t)-T_i(t)), \\
	&\quad \quad t \geq 0, i=1,2,\cdots, N.
	\end{aligned}
	\right.
	\end{equation}
We refer to the model as the \emph{generic simplified URQT (SURQT) model}. See Fig. 4 for the state transition rates of person $i$ under the model.

\begin{figure}[!t]
	\centering
	\includegraphics[width=2.5in]{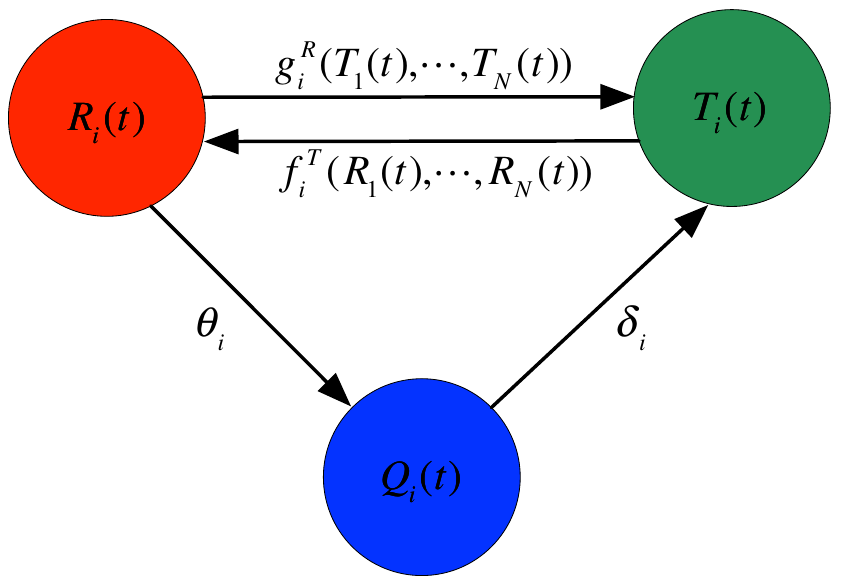}
	\caption{The state transition rates of person $i$ under the generic SURQT model.}
\end{figure}

According to the asymptotically autonomous system theory, the asymptotic dynamics of the generic SURQT model is quantitively similar to that of the generic URQT model. Therefore, the focus of the next section is on the asymptotic dynamics of the generic SURQT model.

Let
\[
\Omega=\left\{(x_1,\cdots,x_{2N})\in \mathbb{R}_+^{2N}\mid x_i+x_{N+i}\leq 1, 1 \leq i \leq N\right\}.\]
The initial state of model (6) lies in $\Omega$. It is easily shown that $\Omega$ is positively invariant for model (6).

Introduce the following notations.
\[
	\begin{split}
	& \mathbf{R}(t) = (R_1(t), \cdots, R_N(t))^T, \quad \mathbf{T}(t) = (T_1(t), \cdots, T_N(t))^T,\\
	& diag\mathbf{R}(t) = diag\left(R_i(t)\right), \quad diag\mathbf{T}(t) = diag\left(T_i(t)\right), \\
	& \mathbf{f}(\mathbf{R}(t)) = \left(f_1^T(\mathbf{R}(t)), \cdots, f_N^T(\mathbf{R}(t))\right)^T,\quad \mathbf{g}(\mathbf{T}(t))= \left(g_1^R(\mathbf{T}(t)), \cdots, g_N^R(\mathbf{T}(t))\right)^T.
	\end{split}
	\]
Then model (6) can be written as
	\begin{equation}
	\left\{
	\begin{aligned}
	\frac{d\mathbf{R}(t)}{dt}&=diag\mathbf{T}(t)\mathbf{f}(\mathbf{R}(t))- diag\mathbf{R}(t)\mathbf{g}(\mathbf{T}(t))-\mathbf{D}_{\theta}\mathbf{R}(t),\\
	\frac{d\mathbf{T}(t)}{dt}&=diag\mathbf{R}(t)\mathbf{g}(\mathbf{T}(t))-diag\mathbf{T}(t)\mathbf{f}(\mathbf{R}(t))+\mathbf{D}_{\delta}[\mathbf{1}-\mathbf{R}(t)-\mathbf{T}(t)],
	\end{aligned}
	\right.
	\end{equation}
where $\mathbf{1}$ stands for the all-one column vector of dimension $N$.


\section{Dynamics of the generic URQT model}

In this section, we shall investigate the dynamics of the generic SURQT model, which implies the dynamics of the generic URQT model. First, the preliminary knowledges that will be used later are presented below.

\subsection{Preliminaries}

For fundamental knowledge on matrix theory, see Ref. \cite{Horn2013}. In what follows, we consider only real square matrices. Given a matrix $\mathbf{A}$, let $s(\mathbf{A})$ denote the maximum real part of an eigenvalue of $\mathbf{A}$, and let $\rho(\mathbf{A})$ denote the spectral radius of $\mathbf{A}$, i.e., the maximum modulus of an eigenvalue of $\mathbf{A}$. $\mathbf{A}$ is \emph{Metzler} if its off-diagonal entries are all nonnegative.

\begin{lm}(Corollary 8.1.30 in \cite{Horn2013})
	Let $\mathbf{A}$ be a nonnegative matrix. If $\mathbf{A}$ has a positive eigenvector $\mathbf{x}$, then $\rho(\mathbf{A})$ is an eigenvalue of $\mathbf{A}$, and $\mathbf{x}$ belongs to $\rho(\mathbf{A})$.
\end{lm}

\begin{lm}(Lemma 2.3 in \cite{Varga2000})
	Let $\mathbf{A}$ be an irreducible Metzler matrix. Then $s(\mathbf{A})$ is a simple eigenvalue of $\mathbf{A}$, and, up to scalar multiple, $\mathbf{A}$ has a unique positive eigenvector $\mathbf{x}$ belonging to $s(\mathbf{A})$.
\end{lm}

A matrix $\mathbf{A}$ is \emph{Hurwitz stable} or simply \emph{Hurwitz} if its eigenvalues all have negative real parts, i.e., $s(\mathbf{A}) < 0$. A matrix $\mathbf{A}$ is \emph{diagonally stable} if there is a positive definite diagonal matrix $\mathbf{D}$ such that the matrix
$
\mathbf{A}^T\mathbf{D}+\mathbf{D}\mathbf{A}
$
is negative definite.

\begin{lm}(Section 2 in \cite{Narendra2010})
	A Metzler matrix is diagonally stable if it is Hurwitz.
\end{lm}

For fundamental theory on differential dynamical systems, see Ref. \cite{Khalil2002}.

\begin{lm} (Chaplygin Lemma, see Theorem 31.4 in \cite{Szarski1965}) Consider a smooth $n$-dimensional system of differential equations,
\[
\frac{d\mathbf{x}(t)}{dt} = \mathbf{f}(\mathbf(\mathbf{x}(t)), \quad t \geq 0,
\]
and the two associated system of differential inequalities,
\[
\frac{d\mathbf{y}(t)}{dt} \leq \mathbf{f}(\mathbf(\mathbf{y}(t)), \quad t \geq 0
\]
and
\[
\frac{d\mathbf{z}(t)}{dt} \geq \mathbf{f}(\mathbf(\mathbf{z}(t)), \quad t \geq 0,
\]
with $\mathbf{y}(0) = \mathbf{z}(0) = \mathbf{x}(0)$. Suppose that for $i = 1, \cdots, n$ and any $a_1, \cdots, a_n \geq 0$, there holds
\[
f_i(x_1+a_1, \cdots, x_{i-1}+a_{i-1}, x_i, x_{i+1} + a_{i+1}, \cdots, x_n + a_n) \geq f_i(x_1, \cdots, x_n).
\]
Then $\mathbf{y}(t) \leq \mathbf{x}(t), \mathbf{z}(t) \geq \mathbf{x}(t), t \geq 0$.
\end{lm}

\begin{lm} (Strauss-Yorke Theorem, see Corollary 3.3 in \cite{Strauss1967}) Consider a differential dynamical system
	\[
	\frac{d\mathbf{x}(t)}{dt} = \mathbf{f}(\mathbf(\mathbf{x}(t)) + \mathbf{g}(t, \mathbf{x}(t)), \quad t \geq 0,
	\]
	with $\mathbf{g}(t, \mathbf{x}(t)) \rightarrow \mathbf{0}$ when $t \rightarrow \infty$. Let
	\[
	\frac{d\mathbf{y}(t)}{dt} = \mathbf{f}(\mathbf(\mathbf{y}(t)), \quad t \geq 0
	\]
	denote the limit system of this system. If the origin is a global attractor for the limit system, and every solution to the original system is bounded on $[0, \infty)$, then the origin is also a global attractor for the original system.
\end{lm}

For fundamental knowledge on fixed point theory, see Ref. \cite{Agarwal2001}.

\begin{lm} (Brouwer Fixed Point Theorem, see Theorem 4.10 in \cite{Agarwal2001}) Let $C \subset \mathbb{R}^n$ be nonempty, bounded, closed, and convex. Let $f: C \rightarrow C$ be a continuous function. Then $f$ has a fixed point.
\end{lm}

\subsection{A fundamental result}

Consider the generic SURQT model (6). Let $R(t)$ and $T(t)$ denote the fraction at time $t$ of rumor-spreaders and truth-believers in the OSN, respectively. That is,
\[
R(t) = \frac{1}{N}\sum_{i=1}^N R_i(t), \quad T(t) = \frac{1}{N}\sum_{i=1}^N T_i(t).
\]
The main aim of this work is to determine the evolving tendency of $R(t)$ and $T(t)$ over time. We start by bounding $R(t)$ and $T(t)$.

\begin{thm}
Consider model (6). Then
\[
	\limsup_{t\rightarrow\infty}R_i(t)\leq \frac{f_i^T(\mathbf{1})}{f_i^T(\mathbf{1})+\theta_i}, \quad \liminf_{t\rightarrow\infty}T_i(t)\geq \frac{\theta_i\delta_i}{\left[f_i^T(\mathbf{1})+\theta_i\right]\left[f_i^T(\mathbf{1})+\delta_i\right]}.
\]
\end{thm}

The proof of the theorem is left to Appendix B. This theorem manifests that, due to the quarantine and education, nobody in the OSN would believe the rumor forever, and every person in the OSN would believe the truth with a positive probability.

\subsection{The equilibria}

The first step to understanding the dynamics of a differential dynamical system is to examine all of its equilibria. The generic SURQT model might admit two different types of equilibria, which are defined as follows.

\begin{de}
	Let $\mathbf{E} = (R_1, \cdots, R_N, T_1, \cdots, T_N)^T$ be an equilibrium of the generic SURQT model (6). Let $\mathbf{R} = (R_1, \cdots, R_N)^T$, $\mathbf{T} = (T_1, \cdots, T_N)^T$.
	
	\begin{enumerate}
		
		\item[(a)] $\mathbf{E}$ is \emph{rumor-free} if  $\mathbf{R} =\mathbf{0}$. A rumor-free equilibrium stands for a steady OSN state in which there is almost surely no rumor-spreader.
		
		\item[(b)] $\mathbf{E}$ is \emph{rumor} if $\mathbf{R} \neq \mathbf{0}$. A rumor equilibrium stands for a steady OSN state in which there is a rumor-spreader with positive probability.
		
	\end{enumerate}
	
\end{de}

Obviously, the generic SURQT model always admits a unique rumor-free equilibrium $\mathbf{E}_0=(\mathbf{0}^T, \mathbf{1}^T)^T$. For the purpose of examining rumor equilibria of the model, define a Metzler matrix as follows.
\begin{equation}
\mathbf{Q}_1=\frac{\partial \mathbf{f}(\mathbf{0})}{\partial \mathbf{x}}-\mathbf{D}_{\theta},
\end{equation}
where $\frac{\partial \mathbf{\mathbf{f}(\mathbf{0})}}{\partial \mathbf{x}}$ stands for the Jacobian matrix of $\mathbf{f}$ evaluated at the origin. As $G_R$ and $G_T$ are strongly connected, $\mathbf{Q}_1$is irreducible.

\begin{thm}
	The generic SURQT model (6) satisfying $s(\mathbf{Q}_1)>0$ admits a rumor equilibrium. Let $(R_1, \cdots, R_N, T_1, \cdots, T_N)^T$ denote the equilibrium, then $0 < R_i < \frac{\delta_i}{\theta_i + \delta_i}, 1 \leq i \leq N$.
\end{thm}

The proof of the theorem is left to Appendix C. This theorem shows that if $s(\mathbf{Q}_1)>0$, the rumor may not die out.

\subsection{Further analysis}

First, we have the following criterion for the global attractivity of the rumor-free equilibrium of the generic URQT model.

\begin{thm}
	Consider model (6). If $s(\mathbf{Q}_1) < 0$, then the rumor-free equilibrium $\mathbf{E}_0$ attracts $\Omega$. As a result, $R(t) \rightarrow 0$, $T(t) \rightarrow 1$ as $t \rightarrow \infty$.
\end{thm}

The proof of the theorem is left to Appendix D. The theorem shows that if $s(\mathbf{Q}_1) < 0$, the rumor would die out definitely.

This theorem has the following useful corollary.

\begin{cor}
	The rumor-free equilibrium $\mathbf{E}_0$ of model (6) attracts $\Omega$ if one of the following conditions is satisfied.
	
	\begin{enumerate}
		
		\item[(a)] $\rho(\mathbf{Q}_1\mathbf{D}_{\theta}^{-1}+\mathbf{E}_N) < 1$.
		
		\item[(b)] $\rho(\mathbf{B}_T\mathbf{D}_{\theta}^{-1}) < 1$.
		
		\item[(c)] $\sum_{i=1}^{N}\beta_{ij}^T < \theta_j$, $j = 1, 2 \cdots, N$.
		
		\item[(d)] $\sum_{j=1}^{N}\frac{\beta_{ij}^T}{\theta_j} < 1$, $i = 1, 2, \cdots, N$.
		
	\end{enumerate}
\end{cor}

The proof of this corollary is left to Appendix E. Theorem 3 and Corollary 1 provide a set of criteria for the dying out of a rumor.

At the end of this section, let us consider the possibility of the long-term existence of the rumor. Define a matrix as follows.
\begin{equation}
\mathbf{Q}_2 = \mathbf{Q}_1 -  diag\left(\frac{\theta_i}{\theta_i + \delta_i}\right)\frac{\partial \mathbf{f}(\mathbf{0})}{\partial \mathbf{x}} - diag\mathbf{g}(\mathbf{1}).
\end{equation}
\begin{thm}
Consider model (6). If $s(\mathbf{Q}_2)>0$, then for any soloution $(\mathbf{R}(t)^T,\mathbf{T}(t)^T)^T$ to the model with $\mathbf{R}(0) >\mathbf{0}$, there is $c > 0$ such that
\[
\liminf_{t\rightarrow\infty}R_i(t)\geq c, \quad 1 \leq i \leq N.
\]
	
\end{thm}

The proof of the theorem is left to Appendix F. This theorem shows that if $s(\mathbf{Q}_2)>0$, the rumor would be persistent.

Although all the results presented in this subsection are derived for the generic SURQT model, they apply to the generic URQT model equally well.

\section{The accuracy of the linear URQT model}

As was mentioned in Section 2, the exact URQT model accurately captures the average dynamics of the rumor-truth interaction. A question arises naturally: under what conditions does the linear URQT model accurately capture this dynamics? This section aims to give an answer to the question through computer simulations.

For the comparison purpose, we need to numerically solve the exact URQT model. Based on the standard Gillespie algorithm for numerically solving continuous-time Markov chain models \cite{Gillespie1977}, we have developed an algorithm for numerically solving the exact URQT model. The basic idea is to take the average of a set of $M$ sample paths of the original URQT model as the solution to the exact URQT model. In our experiments, the number of sample paths in the algorithm is $M = 10^{4}$. As the initial state in the following experiments, a randomly chosen person is set to be rumor-spreading, a randomly chosen person is set to be truth-believing, and all the remaining persons are set to be uncertain.

\subsection{Experiments on scale-free networks}

Scale-free networks are a large class of networks having widespread applications \cite{Albert2002}. Take a randomly generated scale-free network with 100 nodes as the rumor-spreading network as well as the truth-spreading network. By taking random combinations of the parameters, we get 729 pairs of linear and exact URQT models, which are divided into two subsets:  351 pairs for each of which $R(t)$ approaches zero (see Fig. 5 for the comparison results of two pairs in the subset), and 378 pairs for each of which $R(t)$ approaches a nonzero value (see Fig. 6 for the comparison results of two pairs in the subset).

\begin{figure}[H]
   \begin{center}
   \subfigure{\includegraphics[width=0.4\textwidth]{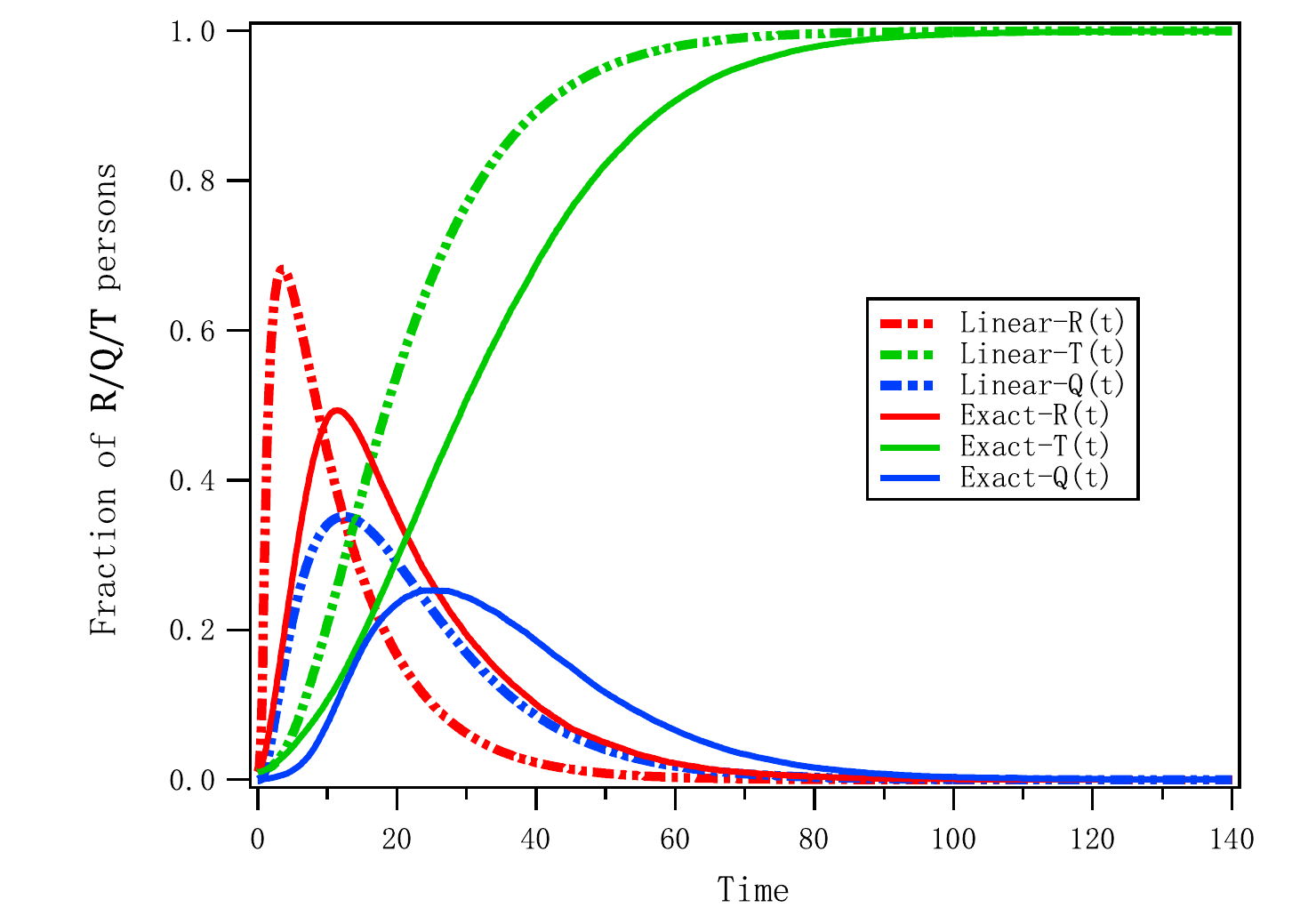}
   \label{fig:a} }
   \subfigure{\includegraphics[width=0.4\textwidth]{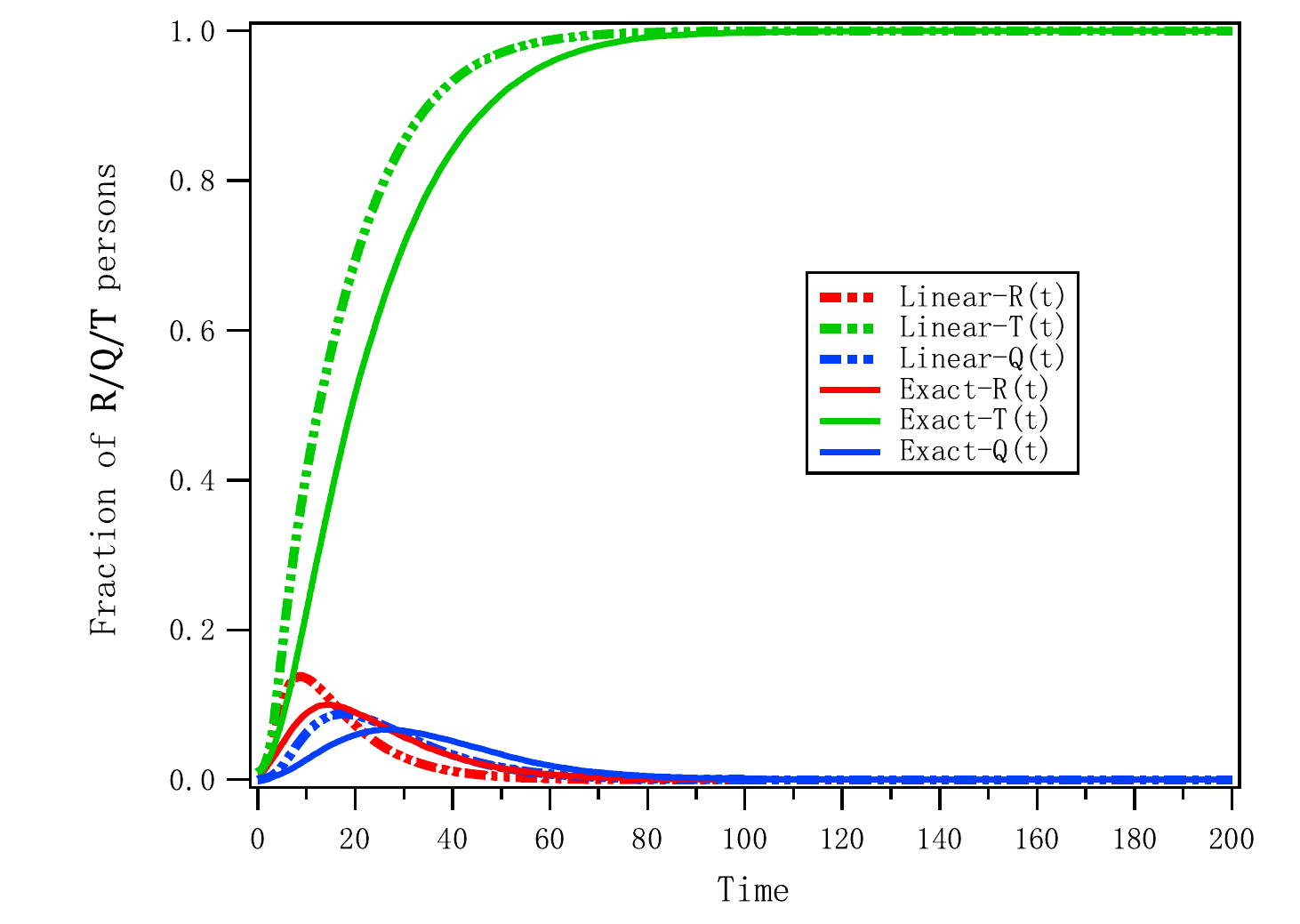}
   \label{fig:b} }
   \vspace{-5ex}
   \hspace{3cm}
   \caption{Comparison results for two pairs in the first collection.}
   \end{center}
\end{figure}
\begin{figure}[H]
   \begin{center}
   \subfigure{\includegraphics[width=0.4\textwidth]{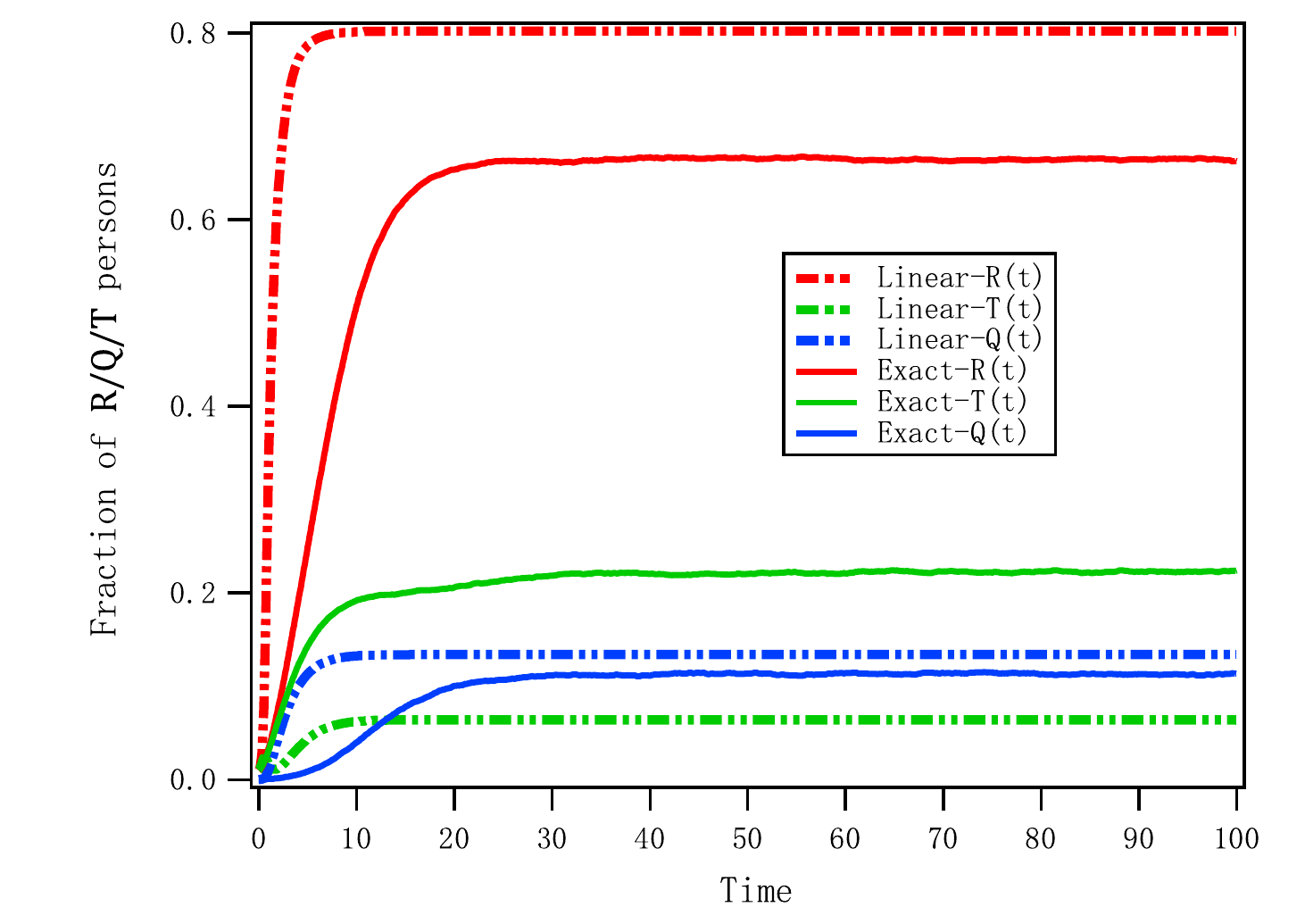}
   \label{fig:a} }
   \subfigure{\includegraphics[width=0.4\textwidth]{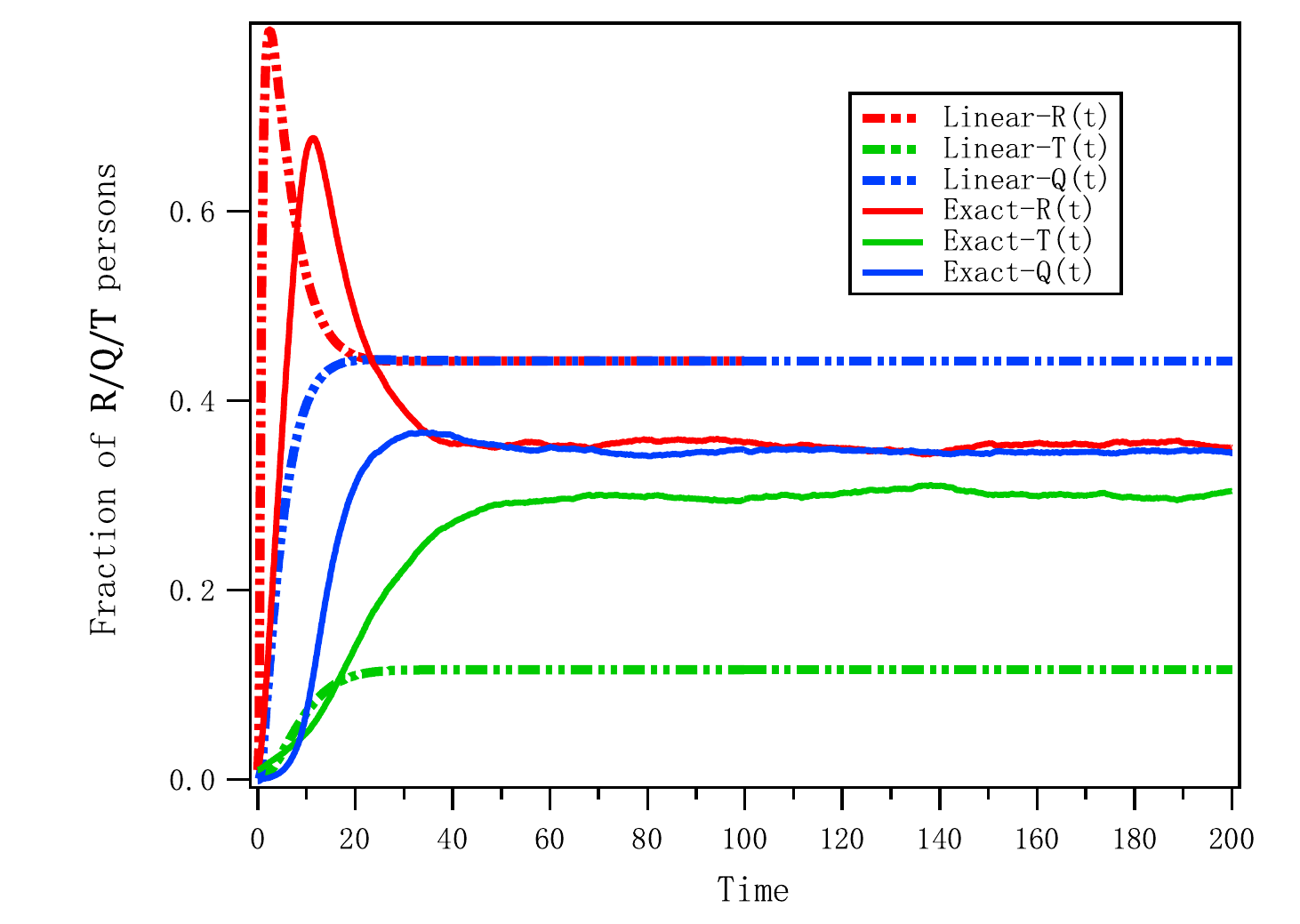}
   \label{fig:b} }
   \vspace{-5ex}
   \hspace{3cm}
   \caption{Comparison results for two pairs in the second collection.}
   \end{center}
\end{figure}

\subsection{Experiments on small-world networks}

Small-world networks are another large class of networks having widespread applications \cite{Watts1998}. Take a randomly generated small-word network with 100 nodes as the rumor-spreading network and the truth-distributing network. By taking random combinations of the parameters, we get 729 pairs of linear and exact URQT models, which are divided into two subsets:  352 pairs for each of which $R(t)$ approaches zero (see Fig. 7 for the comparison results of two pairs in the subset), and 377 pairs for each of which $R(t)$ approaches a nonzero value (see Fig. 8 for the comparison results of two pairs in the subset).

\begin{figure}[H]
   \begin{center}
   \subfigure{\includegraphics[width=0.4\textwidth]{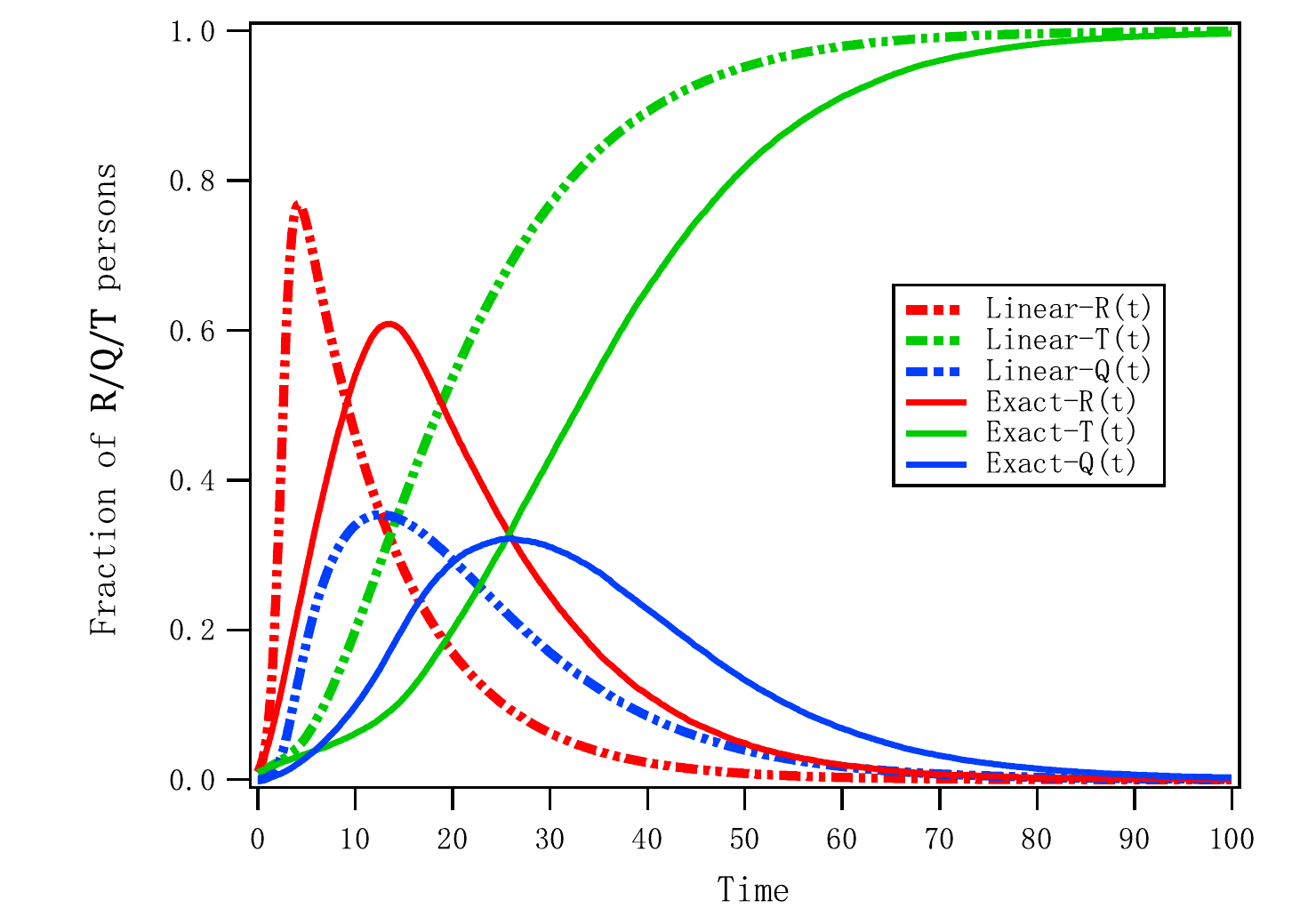}
   \label{fig:a} }
   \subfigure{\includegraphics[width=0.4\textwidth]{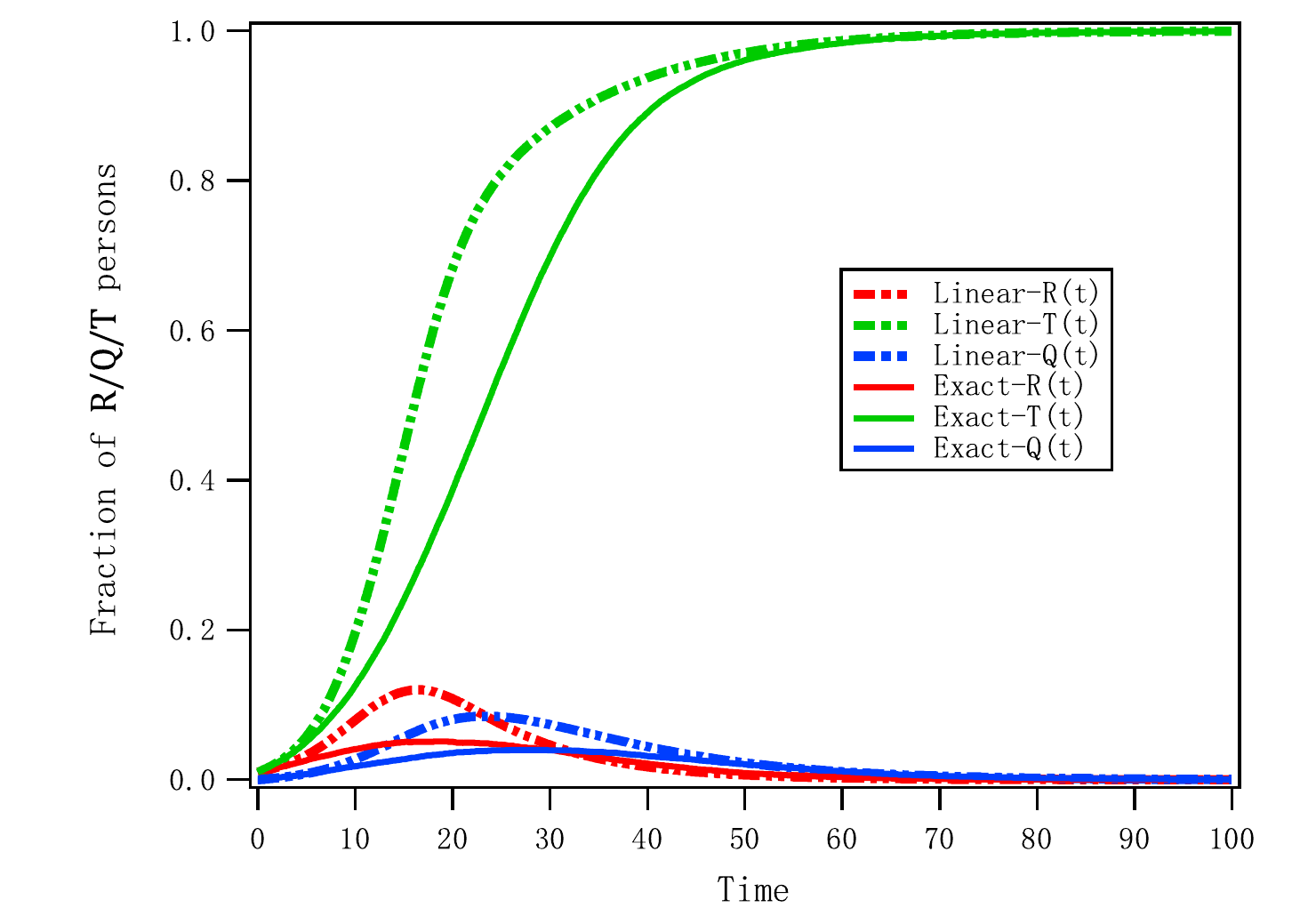}
   \label{fig:b} }
   \vspace{-5ex}
   \hspace{3cm}
   \caption{Comparison results for two pairs in the first subset.}
   \end{center}
\end{figure}
\begin{figure}[H]
   \begin{center}
   \subfigure{\includegraphics[width=0.4\textwidth]{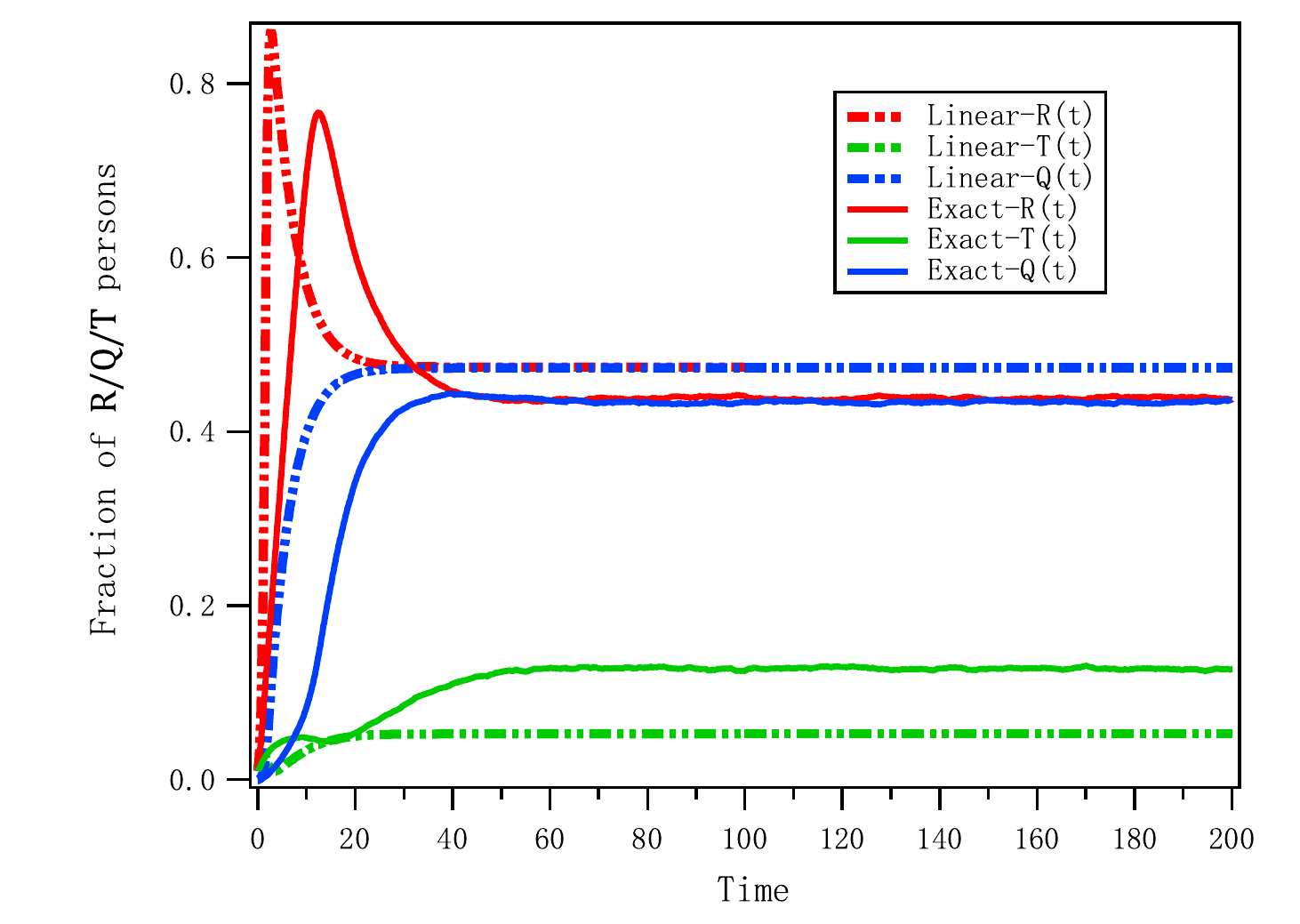}
   \label{fig:a} }
   \subfigure{\includegraphics[width=0.4\textwidth]{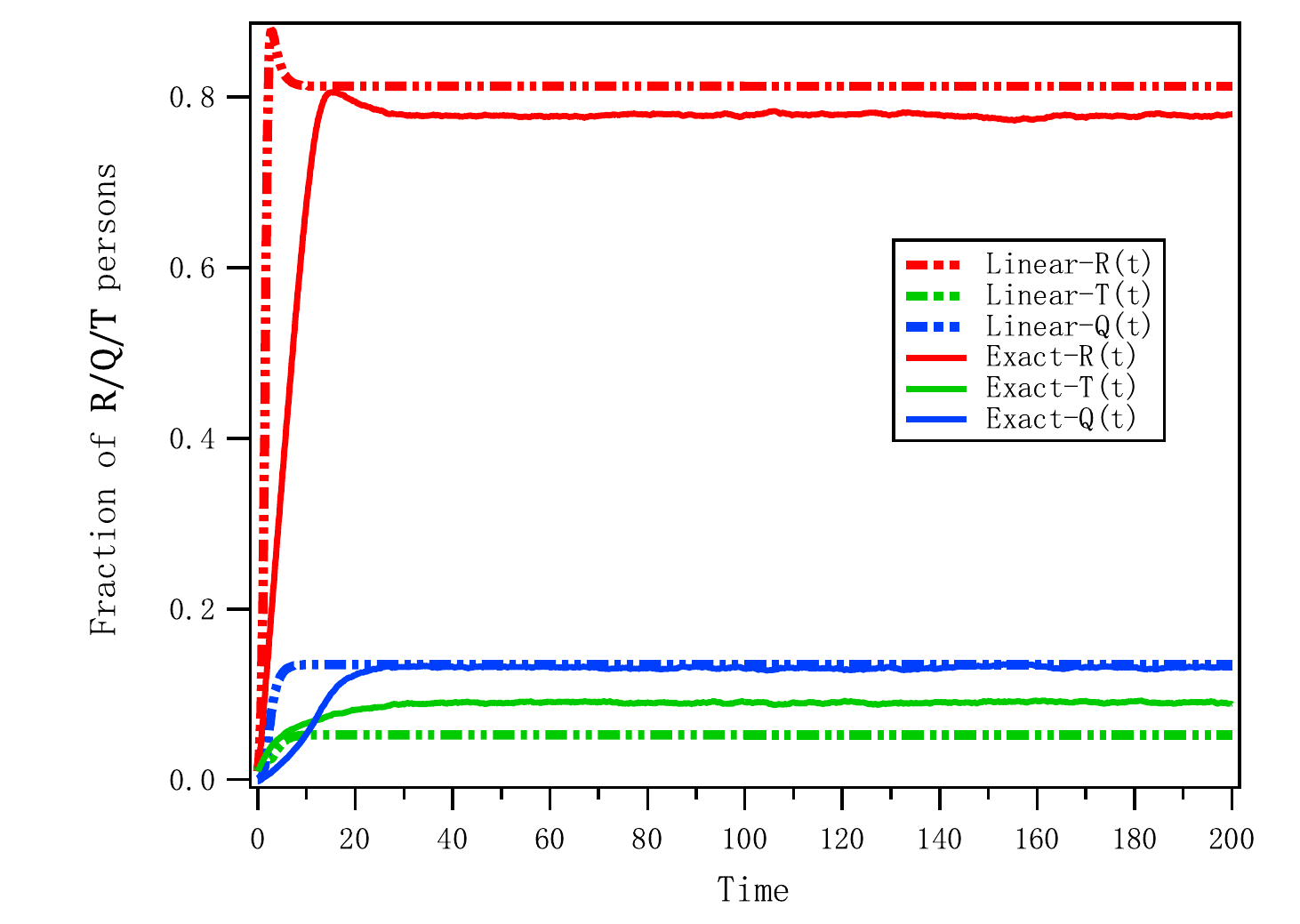}
   \label{fig:b} }
   \vspace{-5ex}
   \hspace{3cm}
   \caption{Comparison results for two pairs in the second subset.}
   \end{center}
\end{figure}

\subsection{A summary}

The following conclusions are drawn from the previous examples.

\begin{enumerate}
	\item[(a)] If $R(t)$ approaches zero, then the linear URQT model satisfactorily captures the average evolution process of the rumor. If $R(t)$ approaches a nonzero value, then the linear URQT model cannot satisfactorily capture the average evolution process of the rumor.
	\item[(b)] The linear URQT model cannot satisfactorily capture the average evolution process of the truth.
\end{enumerate}

In the case where the linear URQT model works well, it can be employed to quickly predict the average evolution dynamics of the rumor in the OSN.

In the case where the linear URQT model doesn't work well, we have to resort to a generic URQT model with nonlinear spreading rates to achieve the goal of accurate prediction. In this case, the idea of deep learning could be employed to accurately estimate the spreading rates \cite{Goodfellow2016}.

\section{Concluding remarks}

This paper has discussed the effectiveness of the truth-spreading/rumor-blocking (TSRB) strategy for inhibiting rumors. A rumor-truth interacting model (the generic URQT model) has been derived. Under the model, a set of criteria for the dying out of a rumor have been given. Extensive simulations show that, when the rumor dies out, the dynamics of a simplified URQT model (the linear URQT model) fits well with the actual rumor-truth interplay process. It is concluded that the generic URQT model (sometimes the linear URQT model) provides a theoretical basis for assessing the effectiveness of the TSRB strategy.

Towards the direction, there are lots of works that are worth study. Under the generic URQT model, the cost paid for restraining a rumor must be minimized \cite{YangLX2016, ZhangTR2017, BiJC2017}. As there is a delay after the appearance of a rumor and before the appearance of the truth, it is necessary to model and study the impact of the delay on the prevalence of the rumor. In the context of individual-level rumor-truth spreading models, it is of practical importance to understand the influence of more factors such as the memory and the mass media on rumor spreading.

\section*{Acknowledgments}

This work is supported by Natural Science Foundation of China (Grant Nos. 61572006, 61379158),
National Sci-Tech Support Plan (Grant No. 2015BAF05B03), Natural Science Foundation of Chongqing (Grant
No. cstc2013jcyjA40011), and Fundamental Research Funds for the Central Universities (Grant No. 106112014CDJZR008823)

\appendix

\section{Proof of Lemma 1}

We prove only the intermediate $N$ equations in this lemma, because the remaining $2N$ equations can be proved analogously. Given a sufficiently small time interval $\Delta t >0$, it follows from the total probability formula that
\begin{equation}
\begin{split}
 R_i(t+\Delta t)&=U_i(t) \Pr\{X_{i}(t+\Delta t)=1 \mid X_{i}(t)=0\}+ R_i(t) \Pr\{X_{i}(t+\Delta t)=1 \mid X_{i}(t)=1\} \\
& \quad + Q_i(t) \Pr\{X_{i}(t+\Delta t)=1 \mid X_{i}(t)=2\}+T_i(t) \Pr\{X_{i}(t+\Delta t)=1 \mid X_{i}(t)=3\}, \quad 1 \leq i \leq N.
\end{split}
\end{equation}
By the conditional total probability formula and in view of model (2), we get that
\[
\begin{split}
&\Pr\{X_{i}(t+\Delta t)=1 \mid X_{i}(t)=0\} \\
&=  \sum_{\mathbf{x} \in \{0, 1, 2, 3\}^N, x_i = 0} \Pr\{X_{i}(t+\Delta t)=1 \mid X_{i}(t)=0, \mathbf{X}(t) = \mathbf{x}\}\cdot \Pr\{\mathbf{X}(t) = \mathbf{x} \mid X_{i}(t)=0\} \\
&= \sum_{\mathbf{x} \in \{0, 1, 2, 3\}^N, x_i = 0} \left[\Delta t \sum_{j = 1}^N \beta_{ij}^U 1_{\{x_j=1\}} + o(\Delta t)\right]\cdot \frac{\Pr\{\mathbf{X}(t) = \mathbf{x}\}}{U_i(t)} \\
&= \frac{\Delta t}{U_i(t)} \cdot \sum_{\mathbf{x} \in \{0, 1, 2, 3\}^N, x_i = 0} \sum_{j = 1}^N \beta_{ij}^U 1_{\{x_j=1\}}\cdot \Pr\{\mathbf{X}(t) = \mathbf{x}\} + o(\Delta t) \\
&= \frac{\Delta t}{U_i(t)} \cdot \sum_{j = 1}^N \beta_{ij}^U \sum_{\mathbf{x} \in \{0, 1, 2, 3\}^N, x_i = 0} 1_{\{x_j=1\}}\cdot \Pr\{\mathbf{X}(t) = \mathbf{x}\} + o(\Delta t) \\
&= \frac{\Delta t}{U_i(t)}  \sum_{j = 1}^N \beta_{ij}^U \Pr\{X_i(t) = 0, X_j(t) = 1\} + o(\Delta t), \quad 1 \leq i \leq N.
\end{split}
\]

\noindent Similarly, we have
\[
\begin{split}
\Pr\{X_{i}(t+\Delta t)=3 \mid X_{i}(t)=1\}
&=\frac{\Delta t}{R_i(t)} \cdot \sum_{j = 1}^N \gamma_{ij}^R \Pr\{X_i(t) = 1, X_j(t) = 3\} + o(\Delta t), \quad 1 \leq i \leq N, \\
\Pr\{X_{i}(t+\Delta t)=2 \mid X_{i}(t)=1\} &= \theta_i \Delta t+o(\Delta t), \quad 1 \leq i \leq N.
\end{split}
\]

\noindent It follows that
\[
\Pr\{X_{i}(t+\Delta t)=1 \mid X_{i}(t)=1\}
=1-\frac{\Delta t}{R_i(t)} \cdot \sum_{j = 1}^N \gamma_{ij}^R \Pr\{X_i(t) = 1, X_j(t) = 3\} -\theta_i \Delta t
 + o(\Delta t), \quad 1 \leq i \leq N.
\]

\noindent Besides, we have
\[
\begin{split}
\Pr\{X_{i}(t+\Delta t)=1 \mid X_{i}(t)=3\}
&= \frac{\Delta t}{T_i(t)} \cdot \sum_{j = 1}^N \beta_{ij}^T \Pr\{X_i(t) = 3, X_j(t) = 1\} + o(\Delta t), \quad 1 \leq i \leq N, \\
\Pr\{X_{i}(t+\Delta t)=1 \mid X_{i}(t)=2\}
&= o(\Delta t), \quad 1 \leq i \leq N.
\end{split}
\]
\noindent Substituting these equations into Eqs. (A.1), rearranging the terms, dividing both sides by $\Delta t$, and letting $\Delta t \rightarrow 0$, we get the second $N$ equations in Lemma 1.
Similarly, we can derive the remaining $2N$ equations in the lemma. The proof is complete.

\section{Proof of Theorem 1}

(a) It follows from the first $N$ equations of model (6) that, for $1\leq i\leq N$,
\[
\frac{dR_i(t)}{dt} \leq f_i^T(\mathbf{1}) - \left[f_i^T(\mathbf{1}) + \theta_i\right]R_i(t), \quad t \geq 0.
\]

\noindent Obviously, the comparison system
\[
\frac{dy_i(t)}{dt} = f_i^T(\mathbf{1}) - \left[f_i^T(\mathbf{1}) + \theta_i\right]y_i(t), \quad t \geq 0
\]

\noindent with $y_i(0)=R_i(0)$ admits $a_i = \frac{f_i^T(\mathbf{1})}{f_i^T(\mathbf{1})+\theta_i}$ as the globally asymptotically stable equilibrium. By Lemma 5, we have
\[
R_i(t)\leq y_i(t), \quad t\geq 0.
\]
So,
\[
\limsup_{t\rightarrow\infty}R_i(t)\leq \lim_{t \rightarrow \infty}y_i(t) = a_i.
\]

(b) Let
\[
b_i = \frac{\theta_i\delta_i}{\left[f_i^T(\mathbf{1})+\theta_i\right]\left[f_i^T(\mathbf{1})+\delta_i\right]}.
\]
Given any sufficiently small $\varepsilon > 0$. It follows from claim (a) of this theorem and the last $N$ equations of model (6) that, for $1\leq i\leq N$, there is $t_i > 0$ such that
\[
\frac{dT_i(t)}{dt} \geq \delta_i(1-a_i-\varepsilon) - \left[f_i^T(\mathbf{1}) + \delta _i\right]T_i(t), \quad t\geq t_i.
\]

\noindent Obviously, the comparison system
\[
\frac{dz_i(t)}{dt} = \delta_i(1-a_i-\varepsilon) - \left[f_i^T(\mathbf{1}) + \delta _i\right]z_i(t), \quad t \geq t_i
\]

\noindent with $z_i(t_i)=T_i(t_i)$ admits $b_i -\frac{\delta_i}{f_i^T(\mathbf{1})+\delta_i}\varepsilon$ as the globally asymptotically stable equilibrium. By Lemma 5, we have
\[
T_i(t) \geq z_i(t), \quad t \geq t_i.
\]
So,
\[
\liminf_{t\rightarrow\infty}T_i(t)\geq \lim_{t \rightarrow \infty}z_i(t) = b_i -\frac{\delta_i}{f_i^T(\mathbf{1})+\delta_i}\varepsilon.
\]
In view of the arbitrariness of $\varepsilon$, we get that $\liminf_{t\rightarrow\infty}T_i(t)\geq b_i$.

\section{Proof of Theorem 2}
Suppose model (6) admits a rumor equilibrium
$\mathbf{E}=(R_1,\cdots,R_N,T_1,\cdots, T_N)^T$. Direct calculations show that
\[
  R_i = \frac{\delta_i(1 - T_i)}{\theta_i + \delta_i} \leq \frac{\delta_i}{\theta_i + \delta_i}, \quad 1 \leq i \leq N.
\]
Suppose some $R_k = 0$. It follows that $f_k^T(R_1,\cdots,R_N)=0$. As $G_R$ is strongly connected, we get that some $\beta_{kl}^T >0$, implying that $R_l = 0$. Repeating this argument, we get that $R_i = 0, 1 \leq i \leq N$, which implies that the equilibrium $\mathbf{E}$ is rumor-free, a contradiction occurs. Hence, $R_i > 0$, $1\leq i \leq N$.

Define a continuous mapping $\mathbf{H}=(H_1,\cdots,H_N)^T: \prod_{i=1}^N (0, \frac{\delta_i}{\theta_i + \delta_i}] \rightarrow (0,1]^N$ by
\[
H_i(\mathbf{x})=\frac{\left(1 - \frac{\theta_i + \delta_i}{\delta_i}x_i\right)f_i^T(\mathbf{x})}{\theta_i+x_ig_i^R(1-\frac{\theta_1 + \delta_1}{\delta_1}x_1,\cdots,1-\frac{\theta_N + \delta_N}{\delta_N}x_N)}, \quad\mathbf{x}=(x_1,\cdots,x_N)^T\in \prod_{i=1}^N (0, \frac{\delta_i}{\theta_i + \delta_i}].
\]
It suffices to show that $\mathbf{H}$ has a fixed point. By Lemma 3, $\mathbf{Q}_1$ has a positive eigenvector $\mathbf{v}=(v_1,\cdots,v_N)^T$ belonging to the eigenvalue $s(\mathbf{Q}_1)$. So,
\[
\mathbf{Q}_1\mathbf{v}=s(\mathbf{Q}_1)\mathbf{v} > \mathbf{0}.
\]
Thus, there is a sufficiently small $\varepsilon > 0$ such that
\[
\begin{split}
  f_i^T(\varepsilon \mathbf{v}) - \theta_i v_i \varepsilon &= \left[\sum_{j=1}^N\frac{\partial f_i^T(\mathbf{1})}{\partial x_j} v_j - \theta_i v_i\right]\varepsilon + o(\varepsilon) \\
  &= s(\mathbf{Q}_1)v_i\varepsilon + o(\varepsilon) \\
  &\geq \varepsilon \frac{\theta_i + \delta_i}{\delta_i}v_i f_i^T(\varepsilon \mathbf{v}) + \varepsilon^2v_i^2g_i^R(1-\frac{\theta_1 + \delta_1}{\delta_1}x_1,\cdots,1-\frac{\theta_N + \delta_N}{\delta_N}x_N) = o(\varepsilon), \quad 1 \leq i \leq N.
\end{split}
\]
That is, $\mathbf{H}(\varepsilon \mathbf{v})\geq \varepsilon \mathbf{v}$. On the other hand, it is easily verified that $\mathbf{H}$ is monotonically increasing, i.e., $\mathbf{u} \geq \mathbf{w}$ implies $\mathbf{H}(\mathbf{u}) \geq \mathbf{H}(\mathbf{w})$.  Define a compact convex set as
\[
K=\prod_{i=1}^N[\varepsilon v_i,\frac{\delta_i}{\theta_i + \delta_i}].
\]
Then $\mathbf{H}|_{K}$
maps $K$ into $K$. It follows from Lemma 7 that $\mathbf{H}$ has a fixed point in $K$. The proof is complete.

\section{Proof of Theorem 3}
Let $(\mathbf{R}(t)^T, \mathbf{T}(t)^T)^T$ be a solution to model (6). It follows from the first $N$ equations of the model that
\[
\frac{d\mathbf{R}(t)}{dt}\leq (\mathbf{E}_N-diag\mathbf{R}(t)) \mathbf{f}^T (\mathbf{R}(t))-\mathbf{D}_{\theta} \mathbf{R}(t), \quad t \geq 0.
\]

Consider the comparison system
\[
\frac{d\mathbf{u}(t)}{dt}=(\mathbf{E}_N-diag\mathbf{u}(t))\mathbf{f}^T(\mathbf{u}(t))-\mathbf{D}_{\theta} \mathbf{u}(t), \quad t \geq 0,
\]
with $\mathbf{u}(0) = \mathbf{R}(0)$. This system can be written as
\[
\frac{d\mathbf{u}(t)}{dt} = \mathbf{Q}_1\mathbf{u}(t) + \mathbf{G}(\mathbf{u}(t)), \quad t \geq 0,
\]
where $\lim_{\mathbf{x} \rightarrow \mathbf{0}}\frac{||\mathbf{G}(\mathbf{x})||}{||\mathbf{x}||} = 0$. Clearly, the comparison system admits the origin as the globally asymptotically stable equilibrium. By Lemma 5, we have
$\mathbf{R}(t)\leq \mathbf{u}(t), t \geq 0$. So, $\mathbf{R}(t) \rightarrow \mathbf{0}$. The limit system of model (6) obtained by letting $\mathbf{R}(t) = \mathbf{0}$ is
\[
  \frac{d\mathbf{v}(t)}{dt} = \mathbf{D}_{\delta}[\mathbf{1} - \mathbf{v}(t)], \quad t \geq 0,
\]
which has $\mathbf{1}$ as the globally asymptotically stable equilibrium. It follows from Lemma 6 that $\mathbf{T}(t)\rightarrow\mathbf{1}$. This completes the proof.

\section{Proof of Corollary 1}
(a) By Theorem 3, it suffices to show $s(\mathbf{Q}_1) < 0$. As $\mathbf{Q}_1\mathbf{D}_{\theta}^{-1}$ is Metzler and irreducible, it follows from Lemma 3 that $\mathbf{Q}_1\mathbf{D}_{\theta}^{-1}$ has a positive eigenvector $\mathbf{x}$ belonging to the eigenvalue $s(\mathbf{Q}_1\mathbf{D}_{\theta}^{-1})$. So,
\[
(\mathbf{Q}_1\mathbf{D}_{\theta}^{-1}+\mathbf{E}_N)\mathbf{x}=[s(\mathbf{Q}_1\mathbf{D}_{\theta}^{-1})+1]\mathbf{x}.
\]
That is, $\mathbf{x}$ is an eigenvector of the nonnegative matrix $\mathbf{Q}_1\mathbf{D}_{\theta}^{-1}+\mathbf{E}_N$ belonging to the eigenvalue $s(\mathbf{Q}_1\mathbf{D}_{\theta}^{-1})+1$. It follows from Lemma 2 and the given condition that
\[
s(\mathbf{Q}_1\mathbf{D}_{\theta}^{-1}) = \rho(\mathbf{Q}_1\mathbf{D}_{\theta}^{-1}+\mathbf{E}_N)-1 < 0.
\]
By Lemma 4, there is a positive definite diagonal matrix $\mathbf{D}$ such that the matrix
\[
\mathbf{P}=(\mathbf{Q}_1\mathbf{D}_{\theta}^{-1})^T\mathbf{D}+\mathbf{D}(\mathbf{Q}_1\mathbf{D}_{\theta}^{-1})
\]
is negative definite. Direct calculations give
\[
\left[\mathbf{D}_{\theta}^{\frac{1}{2}}\mathbf{Q}_1\mathbf{D}_{\theta}^{-\frac{1}{2}}\right]^T\mathbf{D}+\mathbf{D}\left[\mathbf{D}_{\theta}^{\frac{1}{2}}\mathbf{Q}_1\mathbf{D}_{\theta}^{-\frac{1}{2}}\right]
=\mathbf{D}_{\theta}^{\frac{1}{2}}\mathbf{P}\mathbf{D}_{\theta}^{\frac{1}{2}}.
\]
As $\mathbf{D}_{\theta}^{\frac{1}{2}}\mathbf{P}\mathbf{D}_{\theta}^{\frac{1}{2}}$ is negative definite, $\mathbf{D}_{\theta}^{\frac{1}{2}}\mathbf{Q}_1\mathbf{D}_{\theta}^{-\frac{1}{2}}$ is diagonally stable and hence Hurwitz. It follows that
\[
s(\mathbf{Q}_1) = s(\mathbf{D}_{\theta}^{\frac{1}{2}}\mathbf{Q}_1\mathbf{D}_{\theta}^{-\frac{1}{2}}) < 0.
\]

(b) By the concavity of $f_i^T(\mathbf{x})$, we have
$\frac{\partial f_i^T(\mathbf{0})}{\partial x_j}\leq \beta_{ij}^T$. That is, $\mathbf{Q}_1 + \mathbf{D}_{\theta} \leq \mathbf{B}_T$. Hence,
\[
\rho(\mathbf{Q}_1\mathbf{D}_{\theta}^{-1}+\mathbf{E}_N) \leq \rho(\mathbf{B}_T\mathbf{D}_{\theta}^{-1}) < 1.
\]
The claim follows from Claim (a) of this corollary.

(c) The claim follows from Claim (b) of this corollary and $\rho(\mathbf{M}) \leq ||\mathbf{M}||_1$.

(d) The claim follows from Claim (b) of this corollary and $\rho(\mathbf{M}) \leq ||\mathbf{M}||_{\infty}$.

\section{Proof of Theorem 4}
As $s(\mathbf{Q}_2) > 0$, there is $\varepsilon > 0$ such that
\[
  s\left(\mathbf{Q}_2 - \varepsilon\frac{\partial \mathbf{f}(\mathbf{0})}{\partial \mathbf{x}}\right) > 0.
\]
Let $(\mathbf{R}(t)^T, \mathbf{T}(t)^T)^T$ be a solution
to model (6) with $\mathbf{R}(0)\neq \mathbf{0}$. It follows from the model that, for $1\leq i\leq N$,
\[
\frac{dQ_i(t)}{dt}\leq \theta_i - (\theta_i + \delta_i)Q_i(t), \quad t \geq 0.
\]
Clearly, the comparison system
\[
\frac{dq_i(t)}{dt}= \theta_i - (\theta_i + \delta_i)q_i(t), \quad t \geq 0,
\]
with $q_i(0)=Q_i(0)$ admits $\frac{\theta_i}{\theta_i + \delta_i}$ as the globally asymptotically stable equilibrium. So there is $\tau_i>0$ such that
\[
  q_i(t) < \frac{\theta_i}{\theta_i + \delta_i}+\varepsilon, \quad t\geq \tau_i.
\]
By Lemma 5, we have
\[
  Q_i(t) \leq q_i(t) < \frac{\theta_i}{\theta_i + \delta_i}+\varepsilon, \quad t\geq \tau_i.
\]
Let $\tau = \max\{\tau_1,\cdots,\tau_N\}$. It follows from the first $N$ equations of model (6) that
\[
\frac{dR_i(t)}{dt}\geq \left[1-\frac{\theta_i}{\theta_i + \delta_i}-\varepsilon-R_i(t) \right]f_i^T(\mathbf{R}(t)) - g_i^R(\mathbf{1})R_i(t) - \theta_iR_i(t), \quad t \geq \tau, 1\leq i\leq N.
\]
Consider the comparison system
\[
\frac{dw_i(t)}{dt}= \left[1-\frac{\theta_i}{\theta_i + \delta_i}-\varepsilon-w_i(t)\right]f_i^T(\mathbf{w}(t)) - g_i^R(\mathbf{1})w_i(t) -\theta_iw_i(t), \quad t \geq \tau, 1\leq i\leq N.
\]
with $\mathbf{w}(\tau) = \mathbf{R}(\tau)$. It can be written as
\[
\frac{d\mathbf{w}(t)}{dt}= \left[\mathbf{Q}_2 - \varepsilon \frac{\partial \mathbf{f}(\mathbf{0})}{\partial \mathbf{x}}\right]\mathbf{w}(t) + \mathbf{G}(\mathbf{w}(t)), \quad t \geq \tau,
\]
with $\mathbf{w}(\tau) = \mathbf{R}(\tau)$, where $\lim_{\mathbf{x} \rightarrow \mathbf{0}} \frac{||\mathbf{G}(\mathbf{x})||}{||\mathbf{x}||} = 0$. Clearly, there is $c > 0$ such that
\[
  \limsup_{t \rightarrow \infty}||\mathbf{w(t)}|| > c.
\]
By Lemma 5, we have $\mathbf{R}(t) \geq \mathbf{w}(t), t \geq \tau$. Hence,
\[
\limsup_{t \rightarrow \infty}||\mathbf{R}(t)|| \geq \limsup_{t \rightarrow \infty}||\mathbf{w}(t)|| > c.
\]

\end{document}